\documentclass[twocolumn,twoside]{revtex4}
\usepackage{graphicx}
\usepackage{fancyhdr}
\usepackage{amsmath}
\begin{document}

\title{Anomalies in charged-current $B$ decays}

%

\author{Ryoutaro Watanabe}
\affiliation{Instituto Nazionale di Fisica Nucleare,  Sezione di Pisa, 
Largo B. Pontecorvo 3, 56127 Pisa, Italy}

\begin{abstract}
This paper reviews the recent progresses of the flavor and collider searches 
that can probe New Physics effects responsible for the current discrepancy in the lepton flavor universality ratio of $R_{D^{(*)}}$ between the experimental measurements and SM values. 
\end{abstract}

\maketitle
\thispagestyle{fancy}

\section{Introduction}
\label{sec:intro}
The lepton flavor universality ratios of $R_{D^{(*)}} = \Gamma (\bar B \to D^{(*)} \tau\bar\nu) / \Gamma (\bar B \to D^{(*)} \ell\bar\nu)$ have been under the spotlight since BaBar reported its first measurement in 2012. 
Until now, six independent measurements from BaBar~\cite{BaBar:2012obs,BaBar:2013mob}, Belle~\cite{Belle:2015qfa,Belle:2016dyj,Belle:2017ilt,Belle:2019gij,Belle:2019rba}, and LHCb~\cite{LHCb:2015gmp,LHCb:2017smo,LHCb:2017rln}
have presented $R_{D^{(*)}}$ excesses of the experimental data from what is predicted in the Standard Model (SM) as shown in the summary Table~\ref{Tab:ExpSummary}.

\begin{table}[h!]
\begin{center}
\caption{Summary of the experimental $R_{D^\ast}$ and $R_{D}$ measurements, compared with the SM prediction.}
\renewcommand{\arraystretch}{1.5}
\scalebox{0.85}{
\begin{tabular}{rcccc} 
 \hline
 Experiment  &$R_{D^\ast}$ & $R_{D}$ & Correlation & Ref.  \\  
 \hline 
 BaBar (2012)  & ~$0.332(24)(18)$~ & ~$0.440(58)(42)$~ & $-0.31$ & \cite{BaBar:2012obs,BaBar:2013mob} \\
 Belle (2015) & $0.293(38)(15)$ & $0.375(64)(26)$ & $-0.50$ & \cite{Belle:2015qfa} \\
 Belle (2016) & $0.270(35)(28)$ & $-$& $-$ & \cite{Belle:2016dyj,Belle:2017ilt} \\
 Belle (2019) & $0.283(18)(14)$ & $0.307(37)(16)$ & $-0.52$ & \cite{Belle:2019gij,Belle:2019rba} \\
 LHCb (2015) & $0.336(27)(30)$ & $-$ & $-$ & \cite{LHCb:2015gmp} \\
 LHCb (2017) & $0.280(18)(29)$ & $-$ & $-$ & \cite{LHCb:2017smo,LHCb:2017rln} \\
 %
 \hline
 {Average} & {$0.338\pm 0.030$} & {$0.297\pm 0.013$} & $-0.39$ \\ 
 \hline 
 SM (HFLAV) & $0.299 \pm 0.003$ & $0.258 \pm 0.005$ & & \cite{HFLAV:2019otj}  \\
 \hline
\end{tabular}
}
\label{Tab:ExpSummary}
\end{center}
\end{table}

The situation has motivated particle phenomenologists to think about New Physics (NP) interpretations for the excesses in $b \to c \tau \nu$. 
By introducing the NP current 
\begin{equation}
 \label{eq:LX}
 \mathcal L_X = 2\sqrt 2 G_F V_{cb} C_X^\ell (\bar c \, \Gamma b) (\bar \ell \, \Gamma' \nu) \,, 
\end{equation}
it is known that solutions to the world average $R_{D^{(*)}}^\text{WA}$ are given such as 
$C_{\text{VLL}}^\tau \approx 0.09$, $C_{\text{VRL}}^\tau \approx 0.42\,i$, $C_{\text{SLL}}^\tau \approx -0.82 \pm 0.78\,i$, and $C_{T}^\tau \approx 0.15 \pm 0.19\,i$ 
with the currents $(\bar c \gamma^\mu P_L b) (\bar \tau \gamma_\mu P_L \nu)$, $(\bar c \gamma^\mu P_R b) (\bar \tau \gamma_\mu P_L \nu)$, $(\bar c P_L b) (\bar \tau P_L \nu)$, and $(\bar c \sigma^{\mu\nu} P_L b) (\bar \tau \sigma_{\mu\nu} P_L \nu)$, respectively 
in the case of one-operator NP scenarios.

Possible models of a mediator particle for the required $C_X^\tau$ are briefly explained as follows. 
An extra charged vector boson $W'$ could generate $C_{\text{VLL}}$ and $C_{\text{VRL}}$. 
Its UV realization always accompanies an additional neutral vector boson $Z'$ and thus it is usually severely constrained in order to suppress a tree-level FCNC contribution. 
See Ref.~\cite{Kumar:2018kmr} for a comprehensive analysis. 

A charged Higgs boson, realized from an additional Higgs sector, induces $C_{\text{SLL}}$. 
Typical two Higgs doublet models with $Z_2$ symmetry (such as type-II) are not able to produce a large and complex value of $C_{\text{SLL}}$. 
Hence, a general model (sometimes referred to as type-III) is only viable as the solution, which will be discussed later. 

Leptoquarks (LQs) provide alternative solutions to the excesses~\cite{Sakaki:2013bfa,Iguro:2018vqb}. 
Three LQs, $(\bar 3, 1, 1/3)$ scalar $\text{S}_1$, $(3, 2, 7/6)$ scalar $\text{R}_2$, and $(3, 1, 2/3)$ vector $\text{U}_1$ are viable to explain the present anomaly. 
The $\text{S}_1$ LQ induces two independent contributions encoded as $C_{\text{VLL}}$ and $C_{\text{S}_1} \equiv C_{\text{SLL}} = -4C_T\approx 0.13$, where the number in the latter is the solution for this unique combination. 
Similarly, the $\text{R}_2$ LQ has another unique solution with the form of $C_{\text{R}_2} \equiv C_{\text{SLL}} = +4C_T\approx 0.4\,i$. 
Lastly, the $\text{U}_1$ LQ gives $C_{\text{VLL}}$ and $C_{\text{SLL}}$, 
which has been well studied since it could also access another anomaly in neutral-current $B$ decays, which motivated us to propose UV completions as in Refs.~\cite{Calibbi:2017qbu,Heeck:2018ntp,Fornal:2018dqn,Iguro:2021kdw}.

In the next section, recent studies on flavor observables that can probe NP effects for the $R_{D^{(*)}}^\text{WA}$ solutions will be overviewed. 
Next, state-of-the-art collider searches and prospects for the typical NP interpretations, shown above, will be introduced. 
We will also review recent studies for NP possibilities in the light lepton modes and its impact on $R_{D^{(*)}}$.

\section{The light lepton modes}
Before considering NP signals in the tau lepton mode, we show two recent NP studies for the light lepton modes.

In Ref.~\cite{Iguro:2020cpg}, the authors revisited the $|V_{cb}|$ fit analysis from the exclusive process $\bar B \to D^{(*)} \ell\bar\nu$ for $\ell = (e,\mu)$, and extended the analysis including the NP currents as in eq.(\ref{eq:LX}). 
By employing a general form factor (FF) description~\cite{Bordone:2019vic} based on Heavy-Quark-Effective-Theory (HQET), the authors performed simultaneous fit analyses with respect to the FF parameters, $|V_{cb}|$, and $C_X^e = C_X^\mu$. 
Data set for the analyses include such as the full angular distribution data of $\bar B \to D^{*} \ell\bar\nu$ from Belle~\cite{Belle:2018ezy}, and theory evaluations of FF from lattice and LCSR studies. 
See Ref.~\cite{Iguro:2020cpg} for the complete data set and details of the analysis.

 \begin{figure}[h]
 \includegraphics[viewport=0 0 360 354, height=11.5em]{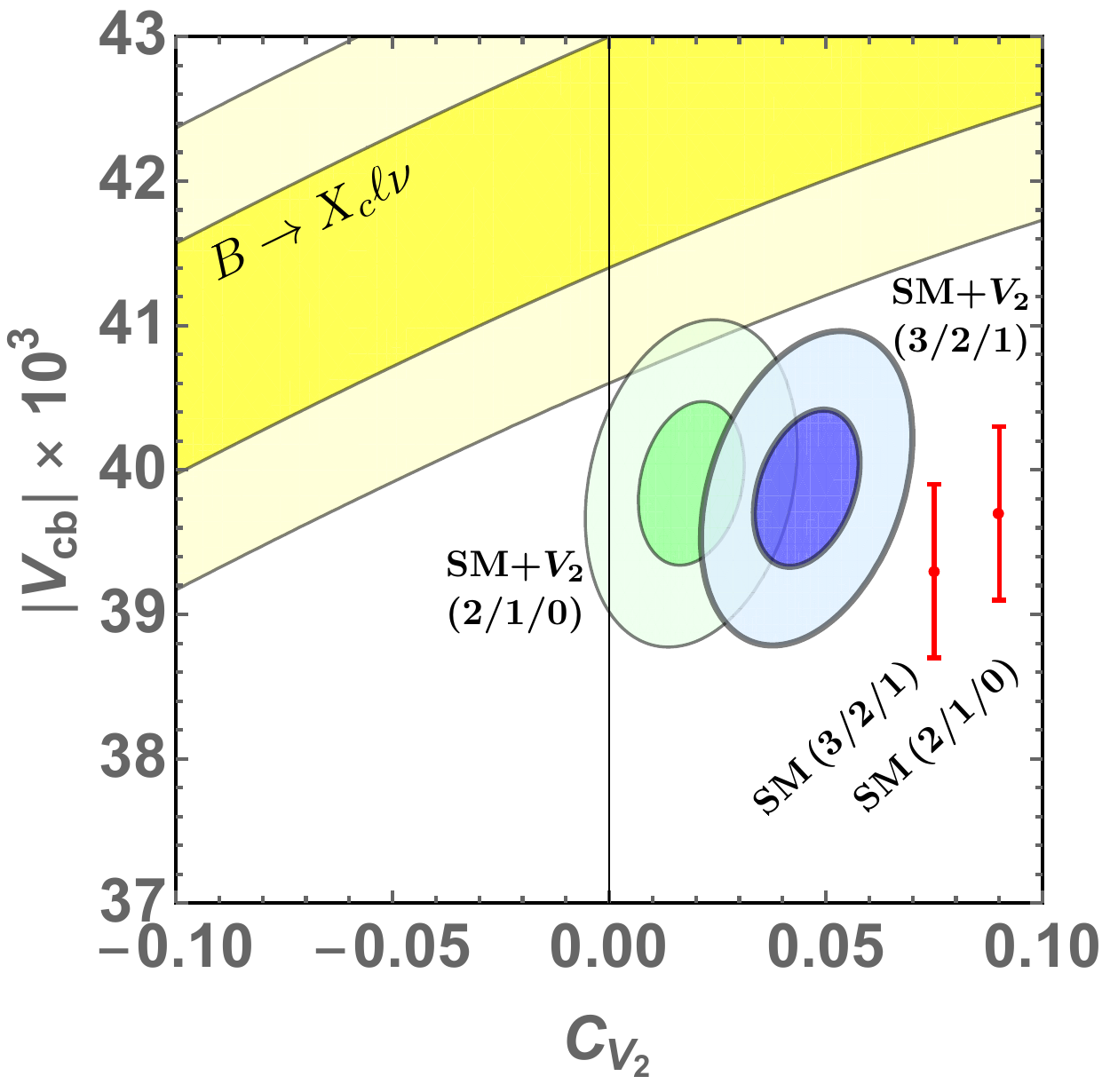}
 \includegraphics[viewport=0 0 360 326, height=11.1em]{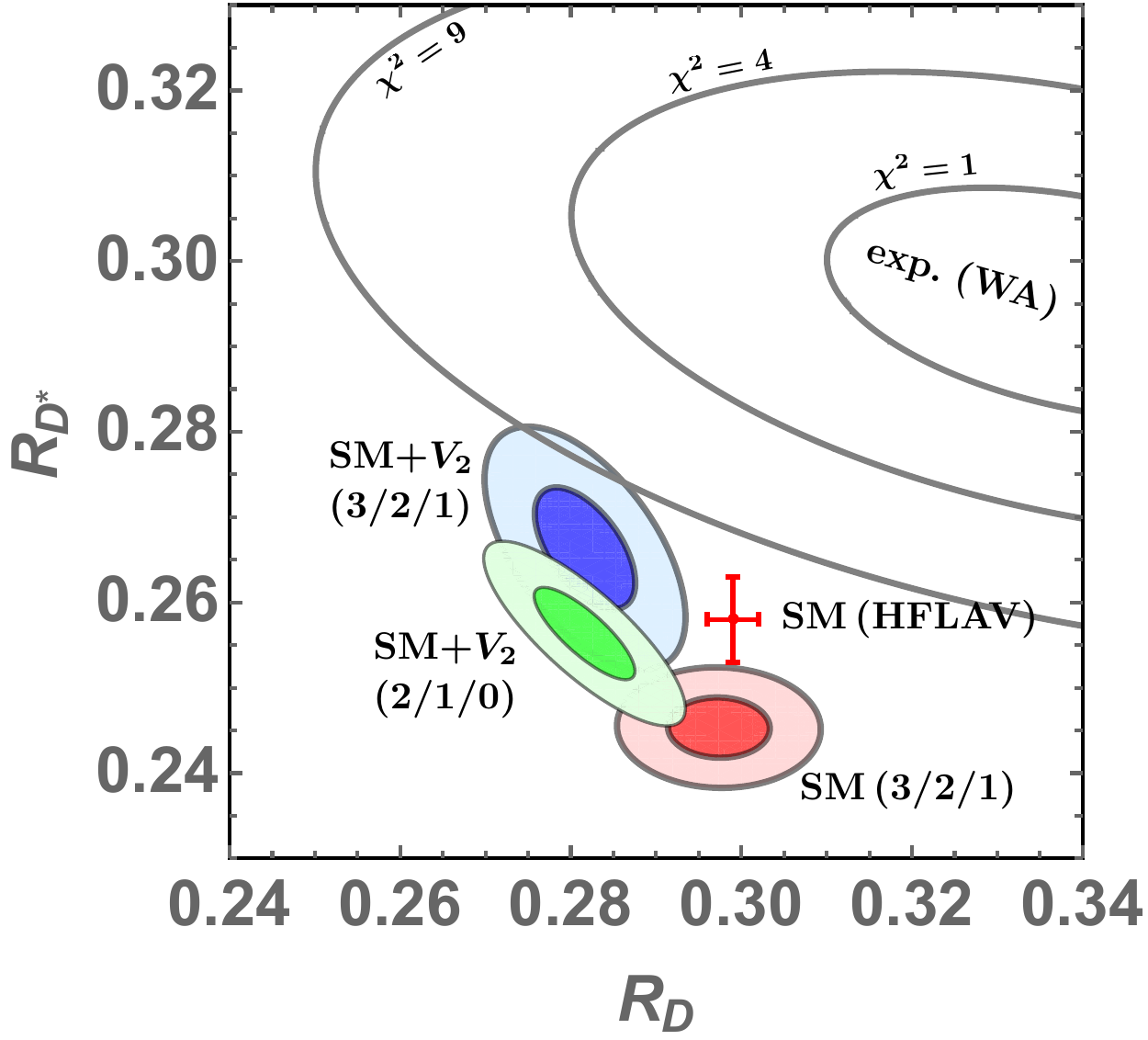}
 \caption{\label{fig:VcbCX} 
 Left:~fitted regions of $C_{\text{VRL}}$ and $|V_{cb}|$ in the SM + VRL scenario. 
 The two distinct FF parameterizations (green and blue) are taken in the fit analysis. 
 See Ref.~\cite{Iguro:2020cpg} for details. 
 %
 Right:~contour plot for $R_D$--$R_{D^*}$ predictions in the $\text{SM} (+ \text{VRL})$ scenarios, where the color correspondence is same as the left panel. 
}
 \end{figure}

One of the fit results for $|V_{cb}|$ and $C_X^\ell$ are shown in the left panel of Fig.~\ref{fig:VcbCX}, where the $\text{SM} + \text{VRL}$ scenario is considered and two different HQET-based FF parameterization models are taken. 
Then, it can be seen that the NP contribution up to $\sim 5\%$ of the SM size ($= 2\sqrt 2 G_F V_{cb}$) can be hidden behind the present $V_{cb}$ measurement. 
The NP presence in the light lepton modes affects $R_{D^{(*)}}$ differently from the usual cases of the aforementioned NP scenarios in the tau mode. 
For the present case, ``NP prediction'' on $R_{D^{(*)}}$ is possible as shown in the right panel of Fig.~\ref{fig:VcbCX}, which indicates that $R_{D*}$ increases while $R_D$ decreases compared with the SM prediction. 
In any case, however, its impact is mild and it is difficult, as easily expected, to accommodate the present $R_{D^{(*)}}$ excesses.

In Ref.~\cite{Bobeth:2021lya}, the authors reconstructed angular observables in $\bar B \to D^{*} \ell\bar\nu$ by utilizing the full angular distribution data from Belle~\cite{Belle:2018ezy}, and found a novel LFU violation in the lepton forward-backward asymmetry, 
defined as $\Delta A_\text{FB} = A_\text{FB}^{(\mu)} - A_\text{FB}^{(e)}$ where we can extract $ A_\text{FB}^{(\ell)}$ from the differential distribution of 
\begin{align}
 \label{eq:AFB}
 {1 \over \Gamma^{(\ell)} } {d\Gamma^{(\ell)} \over d\cos\theta_\ell} 
 = 
 {1 \over 2} + A_\text{FB}^{(\ell)} \cos\theta_\ell + \cdots \,, 
\end{align}
where $\cdots$ is irrelevant for now. 
The SM current gives $\Delta A_\text{FB}^\text{SM} \approx -5.33 \times 10^{-3}$\footnote{
The mass difference $m_e \neq m_\mu$ generates non-zero value in $\Delta A_\text{FB}^\text{SM}$, which is very suppressed though. 
}
while the reconstructed experimental value is obtained as $\Delta A_\text{FB}^\text{exp} \approx +0.0349 (89)$. 
Thus, we can see a $\sim 4\sigma$ tension, which may imply NP existence in the light lepton mode with $C_X^e \neq C_X^\mu$.

NP interpretations have been studied by now in Refs.~\cite{Bobeth:2021lya,Bhattacharya:2022cna}. 
A point of concern is that a NP contribution responsible for $\Delta A_\text{FB}^\text{exp}$ has to satisfy the LFU likely bound of $\mathcal B (\bar B \to D^{*} \mu\bar\nu) / \mathcal B (\bar B \to D^{*} e\bar\nu) = 1.00 \pm 0.01$~\cite{PDG2020}. 
Due to this bound, possible solutions in the literature have no big impact on $R_{D^{(*)}}$.

\section{Flavor Signals}

\subsection{$B_c$ lifetime}
\begin{figure}[t]
 \includegraphics[viewport=0 0 360 360, width=13em]{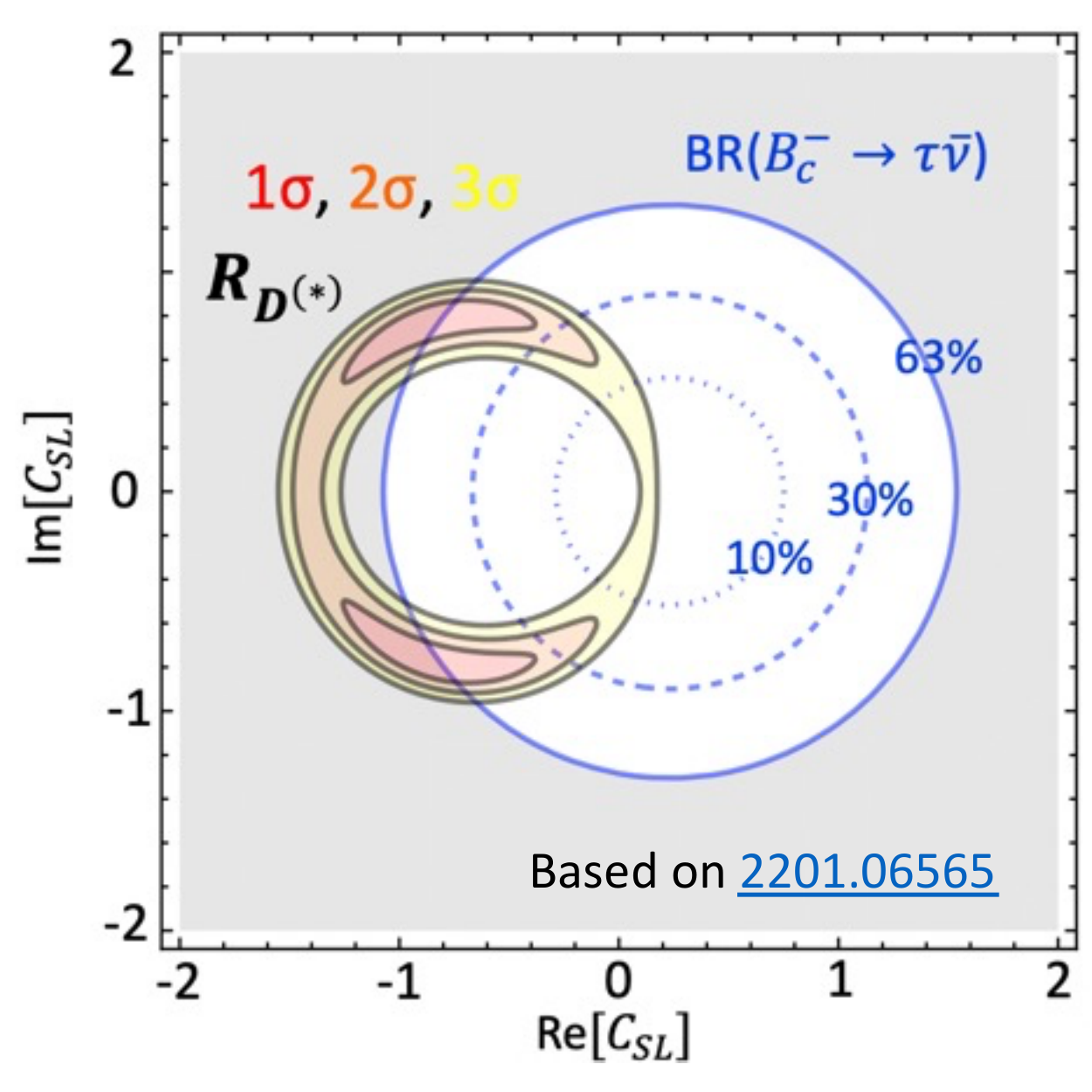}
 \caption{\label{fig:Bclife} 
The allowed region of $C_X^\tau$ from $R_{D^{(*)}}$ (red, orange, yellow) and that from the $B_c$ lifetime (blue lines) with specific conditions on $\mathcal B(B_c \to \tau\nu)$ in the SLL scenario. 
}
\end{figure}

The $B_c$ lifetime has been a good tool to constrain NP that can explain the $R_{D^{(*)}}$ excesses. 
The original idea~\cite{Alonso:2016oyd} was that the lifetime can be evaluated as $0.4\,\text{ps} \lesssim \tau_{B_c}^\text{th} \lesssim 0.7\,\text{ps}$ at $1\sigma$ in the SM~\cite{Beneke:1996xe} 
and hence a decay rate of a NP induced process must not exceed the experimental value $\tau_{B_c}^\text{exp} \approx 0.5 \,\text{ps}$ within the uncertainty. 
The condition is then simplified as $\mathcal B (B_c \to \text{induced by NP}) \lesssim 30\%$ in the original work~\cite{Alonso:2016oyd}. 
For the present case, it corresponds to $B_c \to \tau\nu$. 
This was indeed very crucial for the scalar type NP (SLL) solution to be excluded as can be seen in Fig.~\ref{fig:Bclife}.

However, it has been pointed out~\cite{Blanke:2018yud} that the theory evaluation $\tau_{B_c}^\text{th}$ potentially contains uncertainty from the charm mass input, which was not taken, and thus a more conservative bound should be 
\begin{align}
 \mathcal B(B_c \to \tau\nu) \lesssim 60\% \,. 
\end{align}
This statement has also been supported by the updated study~\cite{Aebischer:2021ilm}. 
Therefore, the SLL solution (probably induced by the charged Higgs or LQ particles) has been revived at present and it is a good candidate to be investigated in detail again. 

\subsection{$B_c$ decay}
Up to now, only a few $B_c$ decay processes have been observed, which is why the $B_c$ lifetime approach is in effect. 
On the other hand, the LFU observable, $R_{J/\psi} = \Gamma (\bar B_c \to J/\psi \tau\bar\nu) \big / \Gamma (\bar B \to J/\psi \mu\bar\nu)$, was measured as $R_{J/\psi}^\text{LHCb} = 0.71 \pm 0.17 \pm 0.18$ for the first time at the LHCb in 2017~\cite{LHCb:2017vlu}. 
Although the present measurement contains a large uncertainty ($35\,\%$), less sensitive to the NP effect, it is estimated that the total uncertainty goes down to $8\,\%$ at the LHCb run~3 in five years. 
Thus, the future experimental reach is sufficiently significant to test the $R_{D^{(*)}}$ excesses. 

The SM prediction of $R_{J/\psi}$ has relied on developments of theory evaluations for the $\bar B_c \to J/\psi$ form factors. 
The first study was based on perturbative QCD in 2017~\cite{Wang:2012lrc}, followed by the LCSR study in 2019~\cite{Leljak:2019eyw}. 
Given the recent lattice study~\cite{Harrison:2020gvo}, we obtain the present value as  
\begin{align}
 R_{J/\psi}^\text{SM} = 0.258 \pm 0.004 \,. 
\end{align}
The NP solutions to the $R_{D^{(*)}}$ excesses give predictions on $R_{J/\psi}$, where a general formula including all the NP currents is obtained in Ref.~\cite{Watanabe:2017mip}. 
For instance, the VLL current strongly correlates $R_{D^{*}}$ with $R_{J/\psi}$ such that the VLL solution predicts $0.28 \leq R_{J/\psi}^\text{VLL} \leq 0.29$. 
Later we will show our correlation plot with other relevant observables. 


\subsection{$\Lambda_b$ decay}
Another LFU observable is available in $R_{\Lambda_c} = \Gamma (\Lambda_b \to \Lambda_c \tau\bar\nu) \big / \Gamma (\Lambda_b \to \Lambda_c \mu\bar\nu)$. 
Very recently, LHCb measured the tau mode and reported $R_{\Lambda_c}^\text{LHCb} = 0.242 \pm 0.026 \pm 0.040 \pm 0.059$ for the first time~\cite{LHCb:2022piu}, where the measurement of the light lepton mode was taken from the result at DELPHI~\cite{DELPHI:2003qft}.

The SM prediction was given as $R_{\Lambda_c}^\text{SM} = 0.324 \pm 0.004$, which looks consistent for now. 
The general formula of $R_{\Lambda_c}$ in the presence of all the possible NP currents has been studied as well as $R_{J/\psi}$ and  $R_{D^{(*)}}$. 
From the general formulae for them, one can find the strong correlation such that 
\begin{align}
 \label{eq:RLsum}
 { R_{\Lambda_c} \over R_{\Lambda_c}^\text{SM} } 
 =
 0.28 {R_{D} \over R_{D}^\textbf{SM}}
 + 
0.72 {R_{D^*} \over R_{D^*}^\textbf{SM}}
 +
 \delta \,, 
\end{align}
where $\delta$ is a function of $C_\text{SLL}^\tau$, $C_\text{SRL}^\tau$ and $C_T^\tau$, but very suppressed as long as $|C_T| \ll 1$, e.g., $\delta \approx -0.03$ for $C_{T}^\tau \approx 0.15 \pm 0.19\,i$.

A key observation is that the world average of $R_{D^{(*)}}^\text{WA}$ as in Table~\ref{Tab:ExpSummary} and the present SM values provide the model-independent fit value of 
\begin{align}
 \label{eq:RLfit}
 R_{\Lambda_c}^\text{fit} =  0.380 \pm 0.013 \pm 0.005 \,. 
\end{align}
This is another indicator to test the $B$ anomaly and implies that the current $R_{\Lambda_c}^\text{LHCb}$ is not consistent with $R_{D^{(*)}}^\text{WA}$ for now. 
Eq.~(\ref{eq:RLsum}) also indicates that measuring $R_{\Lambda_c}$ cannot distinguish a NP type but gives a unique value for every NP solution to $R_{D^{(*)}}^\text{WA}$.

\subsection{NP prediction summary}
\begin{figure}[t]
 \includegraphics[viewport=0 0 370 500, width=24em]{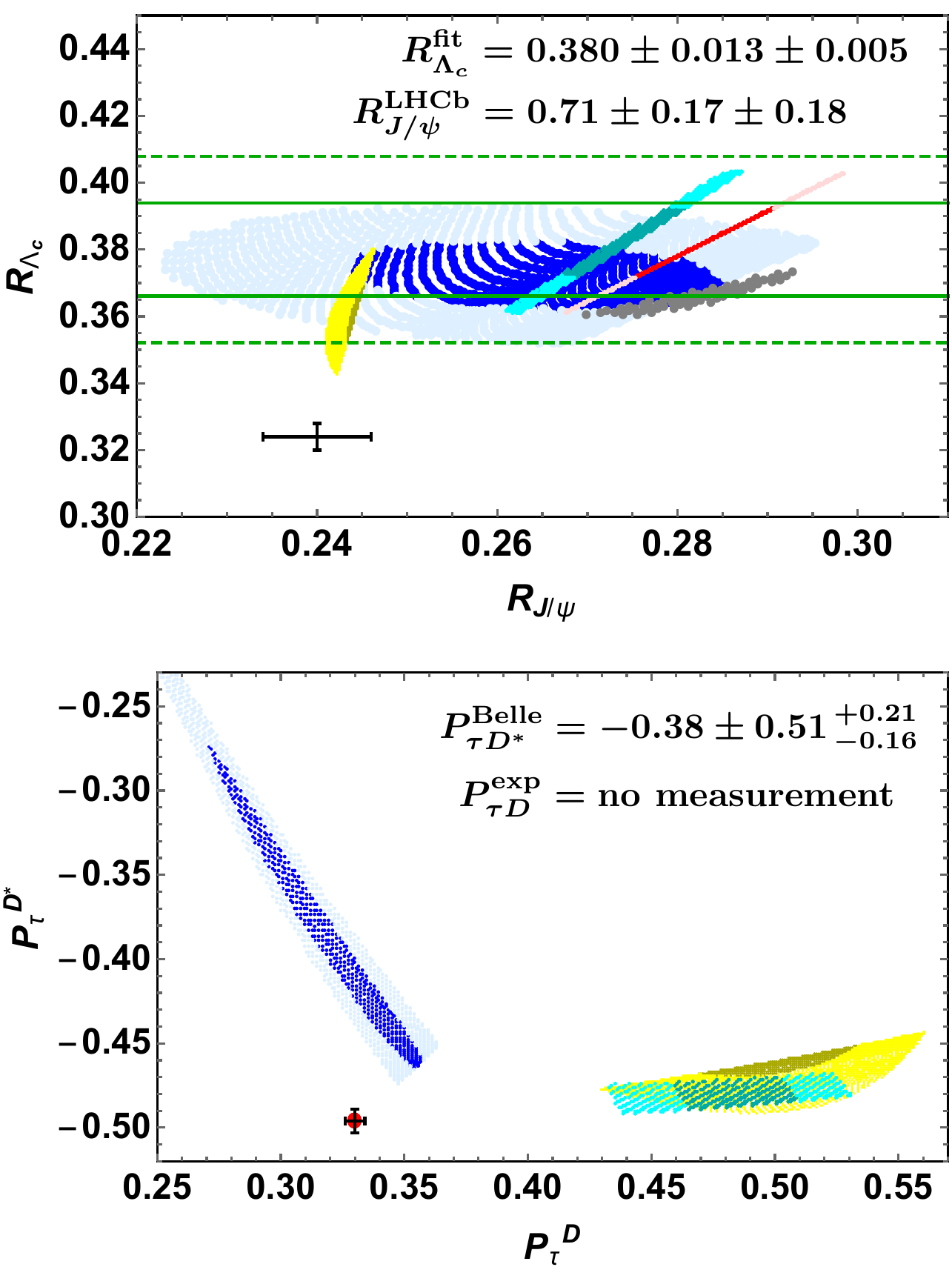}
 \caption{\label{fig:prediction} 
NP predictions on $R_{J/\psi}$, $R_{\Lambda_c}$, and the $\tau$ polarizations $P_\tau^{D^{(*)}}$ in $\bar B \to D^{(*)} \tau\bar\nu$, 
where the NP solutions to $R_{D^{(*)}}^\text{WA}$ are taken within $1\sigma$ and $2\sigma$. 
The black bars are the SM predictions and the color correspondence of the NP scenarios are explained in the main text. 
}
\end{figure}

Assuming the $R_{D^{(*)}}^\text{WA}$ solutions in the NP scenarios, predictions on the observables that can be potentially measured in the future are summarized in Fig.~\ref{fig:prediction}. 
The LFU ratio $R_{J/\psi}$ has a clear correlation with $R_{\Lambda_c}$, where the VLL/VRL/SLL/T/$\text{S}_1$-LQ scenarios are shown in red/gray/yellow/blue/cyan, respectively. 
As argued around Eq.~(\ref{eq:RLfit}), the $R_{\Lambda_c}$ prediction is unique and independent of the NP scenario.

The tau-lepton spin polarizations $P_\tau^{D^{(*)}}$ in $\bar B \to D^{(*)} \tau\bar\nu$ are also known as significant observables, which can be observed at the future $B$-factory experiment. 
See Ref.~\cite{Tanaka:2012nw} for an early stage of this type of work. 
As can be seen in the figure, measuring $P_\tau^{D^{(*)}}$ is effective, particularly, in order to distinguish the T scenario from the others. 

\section{Collider Signals}
The flavor NP current of our concern, $b \to c \tau \nu$, can be investigated at the LHC with the tau and neutrino production process, $b c \to \tau \nu$, which can be searched from tau-lepton + missing events. 
It has been observed as the $W$ boson production in terms of the missing transverse mass $m_T$, 
whose high-mass range, on the other hand, can be used to constrain NPs responsible for the $R_{D^{(*)}}^\text{WA}$ solutions.

The Effective-Field-Theory (EFT) basis study~\cite{Greljo:2018tzh}, adopting the ATLAS~\cite{ATLAS:2018ihk} and CMS~\cite{CMS:2018fza} analyses, shows 
\begin{align}
 |C_\text{VLL}^\text{LHC-EFT}| < 0.32 & \quad\Leftrightarrow\quad C_\text{VLL}^{R_{D^{(*)}}} \approx 0.09 \,, \label{eq:EFTVLL} \\
 |C_\text{VRL}^\text{LHC-EFT}| < 0.33 & \quad\Leftrightarrow\quad C_\text{VRL}^{R_{D^{(*)}}} \approx {0.42\,i} \,, \\
 |C_{T}^\text{LHC-EFT}| < 0.20 & \quad\Leftrightarrow\quad |C_{T}^{R_{D^{(*)}}}| \approx {0.24} \,, \\
 |C_\text{SLL}^\text{LHC-EFT}| < 0.32 & \quad\Leftrightarrow\quad |C_\text{SLL}^{R_{D^{(*)}}}| \approx  {1.13} \,, \label{eq:EFTSLL}
\end{align}
and hence gives a significant impact as one can see that some of the best fit $R_{D^{(*)}}^\text{WA}$ solutions exhibited above (see also Sec.~\ref{sec:intro}) are already excluded. 

\subsection{Leptoquark: $t$-channel process}
In Ref.~\cite{Iguro:2020keo}, it has been pointed out that the EFT results of Eqs.~(\ref{eq:EFTVLL})--(\ref{eq:EFTSLL}) are not the case if the $b c \to \tau \nu$ process is $t$-channel produced such as the LQ scenarios. 
If the NP mass is close to the most sensitive $m_T$ region, the propagator $1/(t - m_\text{NP}^2)$ cannot be approximated as $\simeq -1/m_\text{NP}^2$ that is always appropriate in the case of the flavor observables.
As large $t<0$ generates large $m_T$, it reduces the contribution while the highest sensitivity to the NP bound is fixed at  $m_T \sim 1\,\text{TeV}$ for the present ATLAS and CMS dataset at $36\,\text{fb}^{-1}$. 
This implies that the bound on $C_X$ could be loosen dependent on the NP mediator mass in the case of the $t$-channel process.

This property has been investigated in the literature and the $t$-channel NP bounds for $m_\text{LQ} = 2\text{TeV}/5\text{TeV}$ have been obtained~\cite{Iguro:2020keo} as  
\begin{align}
 &|C_\text{VLL}^\text{LHC-LQ}| < 0.42/0.35 \,,&
 &|C_\text{VRL}^\text{LHC-LQ}| < 0.51/0.38 \,,& \\
 &|C_T^\text{LHC-LQ}| < 0.42/0.24 \,,& 
 &|C_\text{SLL}^\text{LHC-LQ}| < 0.47/0.35 \,,& \\
 &|C_{\text{S}_1}^\text{LHC-LQ}| < 0.56/0.48 \,,&
 &|C_{\text{R}_2}^\text{LHC-LQ}| < 0.53/0.43 \,,& 
\end{align}
which can be compared with Eqs.~(\ref{eq:EFTVLL})--(\ref{eq:EFTSLL}), $|C_{\text{S}_1}^{R_{D^{(*)}}}| \approx 0.13$, and $|C_{\text{R}_2}^{R_{D^{(*)}}}| \approx 0.4$, see also Sec.~\ref{sec:intro}. 
Thus it has been confirmed that the mass dependent bounds are critical for the LQ scenarios since it provides the upper limit of the LQ mediator mass as far as the $R_{D^{(*)}}^\text{WA}$ solution is concerned, which is opposite to the $s$-channel case.

\begin{figure}[t]
 \includegraphics[viewport=0 0 376 210, width=22em]{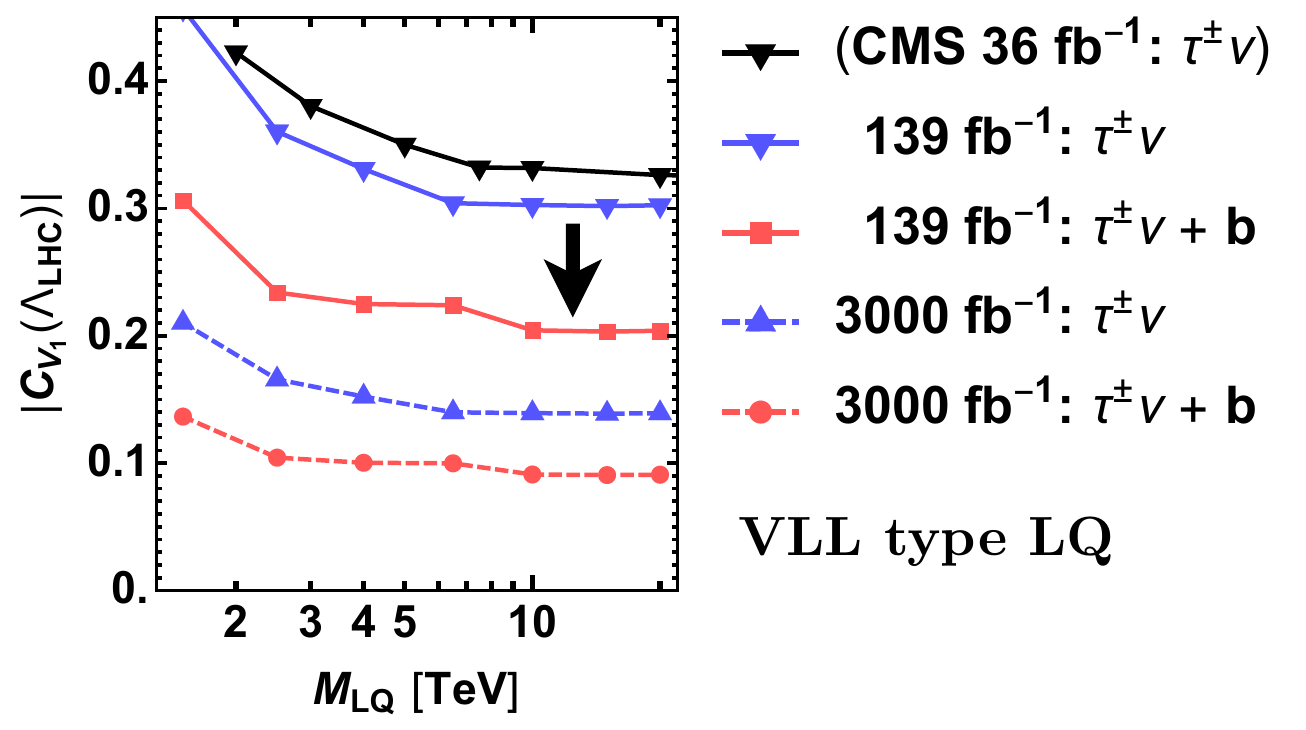}
 \caption{\label{fig:bjet} 
 A prospect of the $C_\text{VLL}$ bound from the $\tau + \text{missing} + b$ search based on the simulation in Ref.~\cite{Endo:2021lhi}. 
 The current bound from  $\tau + \text{missing}$ is also shown for comparison.  
}
\end{figure}

In Ref.~\cite{Endo:2021lhi}, it has been argued that one can improve the search for the $t$-channel NP scenario with an additional $b$-jet tag in the final state, $\tau + \text{missing} + b$. 
The trick is that this $b$-jet requirement greatly reduces the SM background that is generated only by $g q \to b\tau\nu$ for $q=u,c$ and suppressed by $|V_{qb}|^2$, while $t$-channel LQ processes have no such suppression.

This method has been simulated and the prospects for the $C_X$ bound at $139\,\text{fb}^{-1}$ and $3\,\text{ab}^{-1}$ have been obtained. 
Figure~\ref{fig:bjet} shows the simulated result for the LQ with the VLL form together with the current bound from $\tau + \text{missing}$ as shown above. 
See Ref.~\cite{Endo:2021lhi} for all the details of the simulation on the other LQ scenarios.
One can find that requiring  $b$-jet could improve the search potential by $\sim 50\%$.

\subsection{Charged Higgs: loophole}
On the other hand, the EFT result is crucial for the $s$-channel process such as a charged Higgs model as seen in Eq.~(\ref{eq:EFTSLL}). 
Since sensitivity of the present search is gained at $m_T \sim 1\,\text{TeV}$, the charged Higgs mass is constrained up to $m_{H^\pm} \gtrsim 2\,\text{TeV}$. 
However, the range $m_T < 500\,\text{GeV}$ is less sensitive, and thus the low mass region $180\,\text{GeV} \lesssim m_{H^\pm} \lesssim 400\,\text{GeV}$ is a loophole of the exclusion~\cite{Iguro:2022uzz}.

\begin{figure}[t]
 \includegraphics[viewport=0 0 376 210, width=22em]{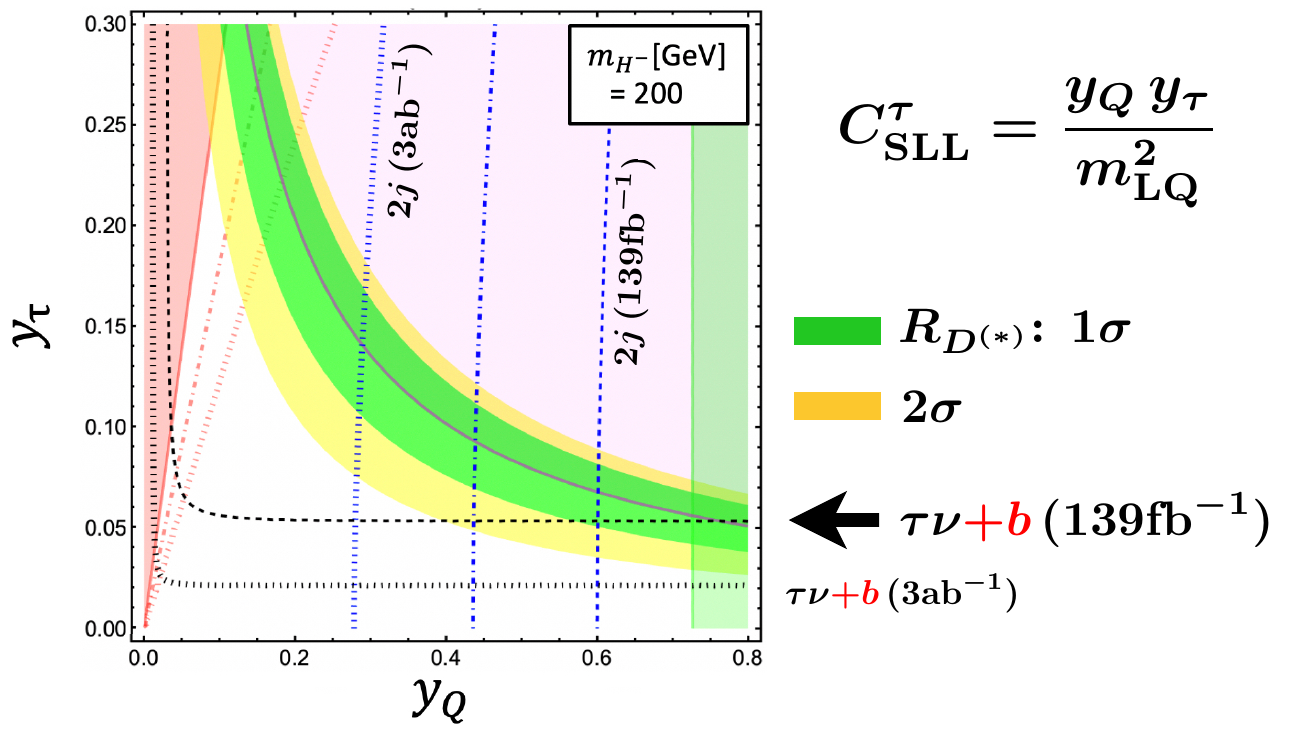}
 \caption{\label{fig:chLHC} 
A prospect for the $\tau \nu + b$ search to constrain the Charged Higgs couplings for the case of $m_{H^\pm} = 200\,\text{GeV}$ at the LHC with $139\,\text{fb}^{-1}$ and $3\,\text{ab}^{-1}$. 
The Charged Higgs solution to $R_{D^{(*)}}^\text{WA}$ is also superimposed. 
}
\end{figure}

In Ref.~\cite{Blanke:2022pjy}, the authors have proposed to investigate this loophole by requiring an additional $b$-jet so that the trigger rate can be suppressed for the final state. 
In Fig.~\ref{fig:chLHC}, the prospect of the LHC bound for the Charged Higgs coupling with $m_{H^\pm} = 200\,\text{GeV}$ is presented, 
and one can see that the present available dataset of $139\,\text{fb}^{-1}$ is sufficient to judge the Charged Higgs scenario as the $R_{D^{(*)}}^\text{WA}$ solution.

\section{Summary}
This paper represents the proceedings of the talk at FPCP 2022 about the recent progresses of the flavor and collider searches 
that can probe the NP effects responsible for the current $R_{D^{(*)}}$ discrepancy between the experiment and SM values.

The solutions for the discrepancy by NPs are encoded in the Wilson coefficients $C_X$. 
Then it has been presented that the flavor signals of these NP scenarios could be seen in the other LFU ratios of $R_{J/\psi}, R_{\Lambda_c}$ and the tau polarizations.

At present, the collider bound from $\tau + \text{missing}$ is crucial such that many of the $R_{D^{(*)}}^\text{WA}$ solutions based on the EFT and $s$-channel type scenarios are already excluded, 
although there exists a loophole of the exclusion in the low mass range of the charged Higgs model. 
This is, however, not the case for NPs of $t$-channel type such as Leptoquark scenarios.

It has also been introduced that a further improvement to constrain (or discover) the NP scenarios is possible 
by searching the $\tau + \text{missing} + b$ event signal for both the $t$-channel (Leptoquarks) and $s$-channel (low mass range of charged Higgs) cases.

\begin{acknowledgments}
This document is adapted from the ``Instruction for producing FPCP2003
proceedings'' by P.~Perret and from eConf templates~\cite{templates-ref}.
\end{acknowledgments}

\end{document}